\documentclass[aps,showpacs,preprint]{revtex4}
\usepackage{epsfig,amssymb,bm,graphics,color}
\newcommand*{\fs}[1]{#1\!\!\!/}
\begin{document}

\title{Neutrino pair emission off electrons in a strong electromagnetic wave field}

\author{A.I.~Titov$^{a,b,c}$, B.~Kampfer$^{a,d}$, H.~Takabe$^{c}$, and A.~Hosaka$^e$}
 \affiliation{
 $^a$Forschungzentrum Dresden-Rossendorf, 01314 Dresden, Germany\\
 $^b$Bogoliubov Laboratory of Theoretical Physics, JINR, Dubna 141980, Russia\\
 $^c$Institute of Laser Engineering, Yamada-oka, Suita, Osaka 565-0871, Japan\\
 $^d$Institut f\"ur Theoretische Physik, TU~Dresden, 01062 Dresden, Germany\\
 $^e$Research Center of Nuclear Physics, 10-1 Mihogaoka Ibaraki, 567-0047 Osaka, Japan}

\begin{abstract}
  The emission of $\nu\bar\nu$ pairs off electrons in a polarized
  ultra-intense  electromagnetic (e.g.\ laser) wave field
  is analyzed. We elaborate on the significance of
  non-linear electrodynamics  effects (i.e., multi-photon processes)
  and the peculiarities of neutrino production.
  Special attention is devoted to the convergence of the reaction probabilities as a function
  of the number of absorbed photons. Expressions for large field intensities are provided.
  The asymmetry between the probabilities of electron and
  $\mu+\tau$ neutrino production
  depends on initial conditions such as energy of the wave field photons and
  the field intensity.
  These findings differ from the lowest order perturbative calculation of the
  reaction $\gamma + e \to e' + \nu \bar \nu$.

\end{abstract}

\pacs{13.35.Bv, 13.40.Ks, 14.60.Ef}
\keywords{Volkov solution, effective low-energy weak interaction, neutrino pair production}

\maketitle

\section{introduction}

The exact solution of Dirac's equation for an electron moving
in the field of a plane electromagnetic wave was found
by D.~M.~Volkov in 1935~\cite{Volkov}.
The electron wave function, compared to the field-free case,
changes due to a modification of its spinor structure and
the appearance of an additional phase factor.
The electron momentum changes to an effective
quasi-momentum, and the electron mass becomes an effective
"dressed" mass. These modifications depend on the
dimensionless variable $\xi^2$ related to the amplitude
of the electromagnetic four-potential $A^\mu$~\cite{LL4}
\begin{eqnarray}
\xi^2=-\frac{e^2\langle A^2\rangle}{M^2_e},
\label{I1}
\end{eqnarray}
where $e$ is the absolute value of electron charge ($e^2=4\pi\alpha$ with
$\alpha \approx 1/137.035$)
and $M_e$ is the electron mass.
(For a manifestly gauge invariant formulation cf.~\cite{Harvei_Heinzl}.)

Decades later, Volkov's solution was applied
to Compton scattering~\cite{Sengupta,Goldman,Brown}
and electron-positron pair production~\cite{Reiss}
in strong electromagnetic fields.
A consistent systematic analysis of these
electromagnetic and further weak processes, such as pion and muon decays,
$\nu\bar \nu$ emission by an electron in an external field etc.\
was performed by Nikishov and Ritus and coworkers in a
series of papers~\cite{NR1,NR2,NR3,NR4} and summarized in the
review \cite{Ritus79}. Later, some aspects of weak interaction,
in particular neutrino emission, by electrons in a strong electromagnetic
field were considered in Refs.~\cite{Lyulka,Merenkov,Skobelev}.
The twofold extension of QED for strong electromagnetic fields
was discussed in the recent paper~\cite{HIM}.

The main result of these previous studies is the conclusion that the
quantum processes are modified significantly in
strong electromagnetic fields. For instance,
an electron can absorb or emit simultaneously a certain number of
field photons, depending on
the initial conditions of the considered process. This fact, together with the
modifications of above mentioned electron properties,
results in strong non-linear and non-perturbative
effects which can not be described within the usual perturbative
quantum electrodynamics (pQED). Consider, for example, the emission
of a photon  with four-momentum $k'$ by an electron moving in a electromagnetic wave field.
The process depends on the invariant variable
$u = \frac{k \cdot k'}{k \cdot p'}$
\cite{LL4},
which varies in the range of $0 < u < u_n = \frac{4 n E_e\omega_L}{M_e^2(1+\xi^2)}$
for the absorption of $n$ photons with four-momenta $k \sim (\omega_L, {\bf k}_L)$
by the electron with four-momenta $p \sim (E_e, {\bf p}_e)$ and $p'$
prior and after the emission process.
One can see that (i) the kinematical limit $u_n$ (phase space)
increases with the number of absorbed photons ("cumulative effect")
and (ii) decreases with increasing field intensity $\xi^2$ because of the electron
mass modification. On the other hand, the contribution of
higher harmonics also increases with $\xi^2$, where,
following~\cite{LL4}, we use the notion
"harmonics" for processes with different $n$'s.

\begin{figure}[h!]
    \includegraphics[width=0.4\columnwidth]{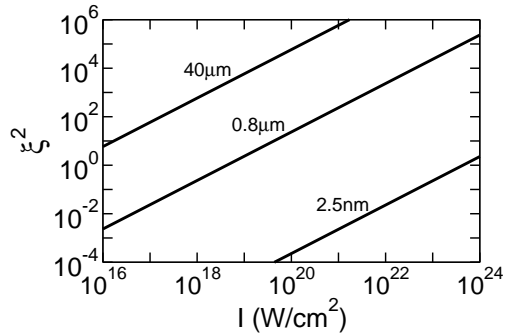}
\caption{\small{
Dependence of $\xi^2$ on the electromagnetic wave field intensity $I$ for
three values of wave length $\lambda_L$.
\label{Fig:1} }}
\end{figure}

Since $\xi^2$ plays an important role, it seems to be
useful to recall the relation between $\xi^2$ and the electromagnetic (laser)
field intensity $I$,
where the electromagnetic field is considered as a classical background field.
For the case of a monochromatic circularly polarized plane wave
with four-potential
$A^\mu=(0,{\bf A})$,
where
${\bf A}(\phi)={\bf a}_x\cos\phi+{\bf a}_y\sin\phi,\,\,\phi=k\cdot x$,
and
$|{\bf a}_x|=|{\bf a}_y|=a,\,\,{\bf a}_x{\bf a}_y=0$,
i.e.\ $k \cdot x=\omega_L t-{\bf k}_L{\bf x}$,
the average value of $A^2$ is equal to $-a^2$, meaning
$\xi^2=\frac{e^2 a^2}{M^2_e}$.
On the other hand, the field intensity may be expressed through the electric ($\bf E$) and
magnetic ($\bf H$) field strengths by
$I=\frac{c}{2}({\bf E}^2+{\bf H}^2)=c{\bf E}^2$.
Taking into account ${\bf E}=-\partial {\bf A}/\partial t$, one gets an expression
for the average intensity $I$ in terms of the amplitude $a$,
$I=ca^2\omega^2$,
which
leads to
$\xi^2
=\frac{\alpha\hbar}{\pi (M_ec^2)^2\,c}\,\lambda_L^2\,I$,
where $\lambda_L=2\pi\hbar c/\omega_L$ is the wave length of the
electromagnetic field $A^\mu$.
The dependence of $\xi^2$ on the electromagnetic field intensity $I$
for different wavelengths $\lambda_L$ is exhibited in Fig.~\ref{Fig:1}.

The wavelength
$\lambda_L = 0.8~\mu$m (or $\omega_L \simeq 1.55$~eV) corresponds to the widely used
titanium-sapphire laser oscillator
(cf.\ Refs.~\cite{C1,C2,C3,C4}). The short wavelength $\lambda_L = 2.5$~nm
(or $\omega_L \simeq 0.5$~keV)
corresponds to the soft x-ray (SXR) free electron laser
at SLAC~\cite{SXR}. The long wavelength $\lambda_L = 40~\mu$m
(or $\omega_L \simeq 0.03$~eV) may be obtained
at the free electron laser for infrared experiments (FELIX)~\cite{FELIX}.
One can see that $\xi^2$ varies within a fairly large range,
depending on the field intensity and wavelength.

In the low-frequency limit, $\omega_L \to 0$, the intensity parameter
becomes large, i.e. $\xi^2\to \infty$ at fixed intensity $I$ or ${\bf E}$.
This limit was considered in some detail by Nikishov and Ritus~\cite{NR1,NR2,Ritus79}
who pointed out that the invariant variable
\begin{eqnarray}
\chi=\frac{e\sqrt{\langle F_{\mu\nu}{p}^\nu\rangle^2}}{M_e^3}
=\xi\frac{k\cdot p}{M_e^2}
\label{I10}
\end{eqnarray}
remains finite and the total probabilities of most of the considered processes
depend only on $\chi$~\cite{Ritus79}.
Here,
$F_{\mu\nu}=\partial_\nu A_\mu -\partial_\mu A_\nu$
is the electromagnetic field tensor.
Such a case of simultaneous limits of
$\xi \to \infty$ and $\omega_L \to 0$ at finite $I$ corresponds to the situation
of an electron interacting with a constant (crossed) electromagnetic field.

Note that two asymptotic regions of the external field were
considered in most of the above quoted papers. One corresponds to the
weak-field limit $\xi^2 \ll 1$. In this case, only
a limited number of harmonics $n \le 2$ contributes.
The opposite case of
large intensity $\xi^2 \to \infty$ with $\omega_L \to0$
allows for two asymptotic limits: $\chi \ll 1$ and $\chi \gg 1$.
Of course, such an analysis of limiting cases is interesting and important by its own.
However, the rapidly evolving laser technology \cite{Tadjima}
can provide conditions where the limit of $\xi^2 \gg 1$ is achieved
at finite $\omega_L$, as well as $\chi \sim 1$ as can be inferred
from Fig.~\ref{Fig:1} and by numerical evaluation
of Eq.~(\ref{I10}).
Therefore, it seems relevant to consider the
probabilities of quantum  processes without the restrictions
imposed in~\cite{Ritus79,Lyulka,Merenkov,Skobelev}.

The goal of present work is accordingly an analysis of neutrino pair
emission off an electron moving in a strong external electromagnetic (laser)
wave field in a wide region of $\xi$ and $\chi$.
Our paper is organized as follows. In Sect.~II,
we consider the neutrino pair emission. A scheme is presented
to overcome convergence problems in the expansion
in terms of harmonics.
The employed method is similar to the one-photon emission
process which is outlined in Appendix A to expose these
similarities and the important differences.
The perturbative neutrino pair emission is recapitulated in Appendix B.
Our conclusions can be found in Sect.~III.

\section{Emission of a neutrino pair \label{IIa}}

\subsection{Basic formulas \label{IIaa}}

\begin{figure}[h!]
  \includegraphics[width=0.25\columnwidth]{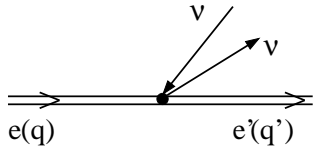}
\caption{\small{
Diagram for the emission of a neutrino pair off an electron
in an external wave field with effective low-energy $e \bar e \nu \bar \nu$ vertex.
The double lines depict electron Volkov states.
\label{Fig:8a} }}
\end{figure}

\begin{figure}[h!]
 \includegraphics[width=0.4\columnwidth]{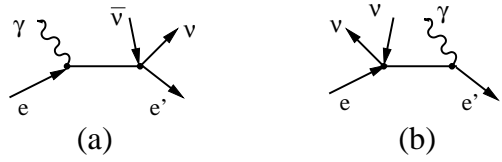}
\caption{\small{
   Lowest order diagrams for the reaction $\gamma + e \to e' + \nu\bar\nu$
   with effective low-energy $e \bar e \nu \bar \nu$ vertices.
   Diagrams (a) and (b) correspond to the
   direct charge and neutral vector boson exchanges.
\label{Fig:8b} }}
\end{figure}

\begin{figure}[h!]
 \includegraphics[width=0.4\columnwidth]{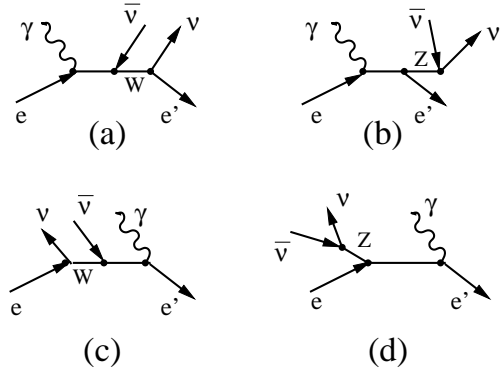}
\caption{\small{
   Lowest order diagram for the reaction $\gamma + e \to e' + \nu\bar\nu$.
   Diagrams (a) and (b) correspond to the
   direct charge (neutral) vector boson exchange;  (c) and (d)
   describe corresponding crossed channels.
\label{Fig:8c} }}
\end{figure}

Similar to the emission of a photon by an electron moving in an
electromagnetic external (background) field (see Appendix \ref{A.1})
one can evaluate the emission of a neutrino pair $\nu_i \bar\nu_i$
of species $i=e,\mu,\tau$ by the $S$ matrix element
\begin{eqnarray}
S^{(i)}_{fi}=-i\frac{G_F}{\sqrt{2}}
\int \bar\psi^*_f (x) \, \gamma^\alpha(C_V^{(i)}-C_A^{(i)} \gamma_5) \, \psi_i (x)
\,L_\alpha^{(i)}
{\rm e}^{iQx}\frac{d^4x}{\sqrt{2E_\nu2E_{\bar\nu}}}
\label{SSS}
\end{eqnarray}
corresponding to the diagram in Fig.~\ref{Fig:8a}.
$M_Q$ is the invariant mass of the $\nu \bar\nu$ pair:
$M_Q^2 \equiv Q^2 = (p_{\nu} +p_{\bar \nu})^2$ with $p_{\nu} \sim (E_\nu, {\bf p}_\nu)$
and $p_{\bar \nu} \sim (E_{\bar \nu}, {\bf p}_{\bar \nu})$
as four-momenta of neutrino and antineutrino.
(We employ units with $\hbar = c = 1$.)
The Volkov solution $\psi_i (x)$ in (\ref{SSS})
(cf.\ Eq.~(\ref{II10}) in Appendix \ref{A.1})
describes the in-state of the electron
in an external wave field, while $\bar\psi^*_f (x)$ refers to the out-state
of the electron also accounting for the external field.
The neutral neutrino current
\begin{eqnarray}
L^{(i)}_\alpha = \bar u_{\nu_i}\gamma_\alpha (1-\gamma_5) v_{\bar\nu_i}~,
\label{III6}
\end{eqnarray}
couples directly to the electron current
$\bar\psi^*_f \, \gamma^\alpha(C_V^{(i)}-C_A^{(i)}\gamma_5) \, \psi_i$
with a strength given by Fermi's constant $G_F=1.66\cdot10^{-5}$~GeV$^{-2}$.
The expression (\ref{SSS}) holds in the local limit where all momenta involved
in the process are much smaller than the masses of the intermediate vector bosons
$Z^0$ and $W^\pm$.
Then, one can obtain "universal" interactions described by the effective low-energy
Lagrangian \cite{Okun} for the direct current-current interaction
\begin{eqnarray}
{\cal L}^{(i)}_{\rm eff}=
\frac{G_F}{\sqrt{2}}
\left[ \bar u_{\nu_i}\gamma_\alpha (1-\gamma_5) v_{\nu_i} \right]
\left[ \bar u_e \gamma^\alpha (C_V^{(i)} - C_A^{(i)}\gamma_5) u_e \right]~,
\label{III3}
\end{eqnarray}
with
\begin{eqnarray}
C_V^{(e)}&=&\frac12+ 2\sin^2\theta_W~,\qquad
C_V^{(\mu,\tau)}=-\frac12+ 2\sin^2\theta_W~, \\
C_A^{(e)}&=&\frac12~, \hspace*{2.9cm} C_A^{(\mu,\tau)}=-\frac12~;
\label{III4}
\end{eqnarray}
the average value of $\sin^2\theta_W\simeq 0.23$ is taken from Ref.~\cite{PDG}.
In the weak-field approximation, where the interaction with the external wave field
is mediated by one photon, the diagram in Fig.~\ref{Fig:8a}
would be resolved by diagrams exhibited in Fig.~\ref{Fig:8b}
corresponding to the process $\gamma + e \to e' + \nu \bar \nu$
which can be dealt with perturbatively (see Appendix \ref{B}).
The effective weak vertices
for the $e \bar e \nu \bar \nu$ interaction, in turn, are resolved in the
tree-level approximation within the standard model as exhibited in Fig.~\ref{Fig:8c}.
The advantage of the diagram in Fig.~\ref{Fig:8a} and Eq.~(\ref{SSS}) is that multi-photon
effects are included which become important at high intensities where the
external wave behaves more and more as a classical field.

As in the case of the emission of one photon (see Appendix \ref{A}),
the differential probability of neutrino pair emission can be represented by
the infinite sum of the partial contributions,
\begin{eqnarray}
dW = \sum\limits_{n=1}^{\infty} dW^{(n)}
\label{III10}
\end{eqnarray}
with
\begin{eqnarray}
dW^{(n)}&=&\frac{G_F^2}{2(8\pi)^3} \,R^{(n)}\frac{du\,dM_Q^2}{q_0(1+u)^2}
\frac{d\phi_{q'}}{2\pi}~, \label{III11_a}\\
R^{(n)}&=&\frac83(Q_\alpha Q_\beta - g_{\alpha\beta} Q^2)
\,{\cal M}^{\alpha\beta}_{n}~,
\label{III11}
\end{eqnarray}
where
$\phi_{q'}$ is the azimuthal angle of the outgoing electron and
\begin{eqnarray}
u=\frac{k \cdot Q}{k \cdot p'}~.
\label{III2}
\end{eqnarray}
Analog to the case of one-photon emission (see Appendix \ref{A.1}) the label $n$
refers to the number of photons absorbed from the external field.
For the sake of simplicity, we skip henceforth the index $i$.
The electron tensor
\begin{eqnarray}
{\cal M}^{\alpha\beta}_{n}=
{\rm Tr}[(\fs p' +M_e)S^\alpha_n(C_V - C_A \gamma_5) (\fs p+M_e)
S^\beta_n(C_V-C_A \gamma_5)]
\label{III12}
\end{eqnarray}
incorporates the electromagnetic wave field via
\begin{eqnarray}
S^\alpha_n=
\left(
\gamma^\alpha-\frac{M_e^2\xi^2k^\alpha\fs k}{2 k\cdot p \, k\cdot p'}
\right)B_n^{(0)}
+\xi M_e\left[\left(\frac{\fs n_1\fs k\gamma^\alpha}{2 k\cdot p'}
+\frac{\gamma^\alpha\fs k\fs n_1}{2 k\cdot p}\right)B^{(1)}_n
\right.
+\left.\left(\frac{\fs n_2\fs k\gamma^\alpha}{2 k\cdot p'}
+\frac{\gamma^\alpha\fs k\fs n_2}{2 k\cdot p}\right)B^{(2)}_n
\right]~.
\nonumber
\end{eqnarray}
Here, ${\bf n}_{1(2)}$ is the unit vector of ${\bf a}_{1(2)}$, and
the functions $B_n$ are related to the Bessel functions
(cf. Refs.~\cite{Ritus79,LL4}):
\begin{eqnarray}
B^{(0)}_n&=&J_n(z){\rm e}^{in\phi_{q'}}~,\nonumber\\
B^{(1)}_n&=&\frac12
\left(J_{n+1}{\rm e}^{i(n+1)\phi_{q'}} + J_{n-1}{\rm e}^{i(n-1)\phi_{q'}}
\right)~,
\nonumber\\
B^{(2)}_n&=&\frac{1}{2i}
\left(J_{n+1}{\rm e}^{i(n+1)\phi_{q'}} - J_{n-1}{\rm e}^{i(n-1)\phi_{q'}}
\right)~.
\label{III13}
\end{eqnarray}
The argument of these Bessel functions is
\begin{eqnarray}
z = \frac{2n\xi}{\sqrt{1+\xi^2}}
\sqrt{
\frac{u}{u_n}
\left(
1-\frac{u}{u_n}
\right) - \frac{1+u}{u_n}\frac{M_Q^2}{(1+\xi^2)M_e^2}}~,
\label{III14}
\end{eqnarray}
where $u_n$ is the kinematical limit of the invariant variable $u$ defined by
\begin{eqnarray}
u_n=\frac{2n(k\cdot p)}{M_e^2(1+\xi^2)}
\label{II17}
\end{eqnarray}
which determines the upper limit of $M_Q^2$ at given $u$ by
\begin{eqnarray}
{M_Q^2}_{\rm max}=M_e^2(1+\xi^2)\frac{u(u_n-u)}{1+u}~.
\end{eqnarray}

The functions $R^{(n)}$ in Eq.~(\ref{III11}) do not depend on $\phi_{q'}$;
the explicit expression reads
\begin{eqnarray}
&&\frac{3}{32}R^{(n)} = C_V^2F_V + C_A^2F_A + 2\lambda C_VC_A F_I ,\\
&&F_V=-M_Q^2(2M_e^2+M_Q^2)J^2_n+\xi^2M_Q^2M_E^2
\left(1+\frac{u^2}{2(1+u)}\right)\Delta J_n^2~, \\
&&F_A=M_Q^2(4M_e^2-M_Q^2)J^2_n+
\xi^2M_E^2\left(M_Q^2+(M_Q^2+2M_e^2)\frac{u^2}{2(1+u)}\right)\Delta J_n^2~,\\
&&F_I=M_Q^2M_e\frac{\xi(2+u)}{\sqrt(1+u)}
\frac{\left(
{M_Q^2}_{\rm max}\left(1-\displaystyle\frac{u_n}{2(u_n-u)}\right) -M_Q^2
\right)}
{\left(
{M_Q^2}_{\rm max} - M_Q^2\right)^{1/2}} J_n\left(J_{n+1}-J_{n-1}\right)~,
\label{III15}
\end{eqnarray}
where $\lambda=\pm1$ is the relative phase (polarization)
of the amplitudes ${\bf a}_1$ and ${\bf a}_2$, and
$\Delta J_n^2=J_{n+1}^2+J_{n-1}^2 -2J_{n}^2$.
Our expression coincides with the result of Ref.~{\cite{Skobelev}}
in a different notation.

\subsection{Numerical evaluation}

\begin{figure}[h!]
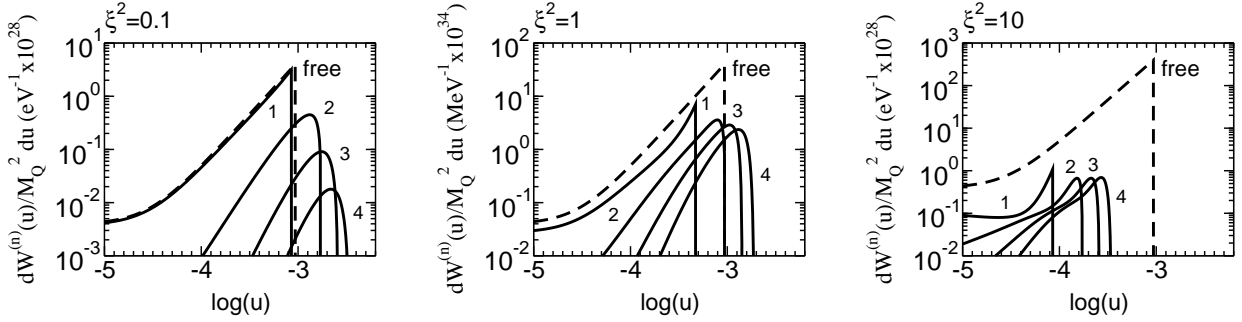

 \includegraphics[width=0.3\columnwidth]{Fig5a.eps}\qquad
 \includegraphics[width=0.3\columnwidth]{Fig5b.eps}\qquad
 \includegraphics[width=0.3\columnwidth]{Fig5c.eps}
   \caption{\small{
   Differential probability of neutrino pair emission as a function of $\log{u}$
   for $\xi^2=0.1,\,1$ and $10$, shown in left, middle and right panels,
   respectively, for the lowest harmonics $n\le4$.
   The result for the reaction $\gamma + e \to e + \nu \bar \nu$
   calculated perturbatively (see Appendix B) is shown by the dashed curves labeled by "free".
   The numbers (from 1 to 4) correspond to the number
   of absorbed photons. The energies of the external (laser) photons
   and incoming electrons are chosen to be $\omega_L = 1.55$~eV and $E_e=40$~MeV,
   respectively. The invariant mass of the outgoing neutrino pair is fixed at $M_Q=10$~eV.
\label{Fig:10} }}
\end{figure}

In Fig.~\ref{Fig:10} we exhibit the differential emission probabilities
summed over all neutrino types for a head-on collision of $40$~MeV electrons
with a laser beam characterized by $\lambda_L = 0.8\,\mu$m (or $\omega_L \simeq 1.55\,$MeV).
We fix the invariant mass of the outgoing neutrino pair by $M_Q=10$~eV.
(The numerical results for the first two harmonics coincide
for $\xi^2=0.1$ with the prediction \cite{Skobelev} based on the asymptotic decomposition
of $W(u)$ for $n=1,2$ in the limit $\xi^2\ll1$.)
The prediction of perturbative QED for the reaction
$\gamma + e \to e' + \nu \bar\nu$ (see Fig.~\ref{Fig:8b} and
Eq.~(\ref{III81})
in Appendix \ref{B}) is shown by the dashed curves.
One can see that for small $\xi^2$ the result for $n=1$ coincides practically
with that of pQED.
Similar to the case of non-linear Compton scattering (see Appendix \ref{A.1}),
even in this case the contribution of higher
harmonics increases the phase space, i.e.\ the kinematic limit, and modifies the total
probability. When $\xi^2$ increases, the difference between the non-perturbative
calculations and pQED is rather large, even for $n=1$,
see middle and right panels of Fig.~\ref{Fig:10}.

\begin{figure}[h!]
 \includegraphics[width=0.5\columnwidth]{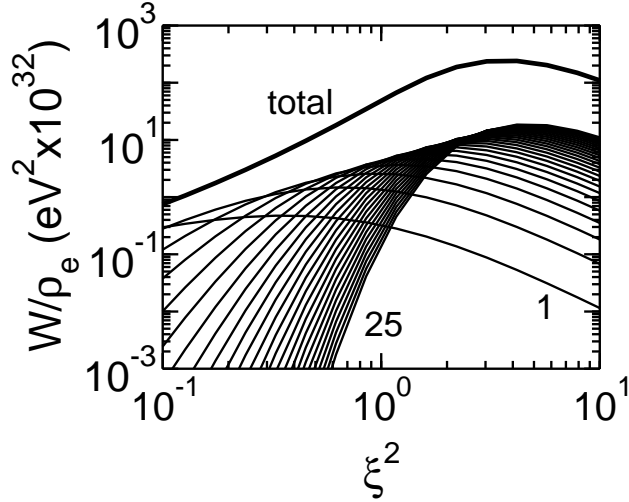}
\caption{\small{
The total probability of neutrino pair emission as
a function of $\xi^2$ for the first 25 harmonics. The thick curve is their total
contribution. The electron and photon energies are chosen as $E_e=40$~MeV and
$\omega_L = 1.55$~eV, respectively.
\label{Fig:11}
}}
\end{figure}

In Fig.~\ref{Fig:11} we show the total probability normalized as
$W/\rho_e$, where $\rho_e$ is the initial electron density
divided by the electron mass in the electron rest frame,
\begin{eqnarray}
\frac{1}{\rho_e}W=\sum\limits_{n=1}^{\infty} \frac{G_F^2}{2(8\pi)^3}
\int\limits_0^{u_{n}}\frac{du}{(1+u)^2}
\int\limits_0^{{M_Q^2}_{\rm max}}\,dM_Q^2\,R^{(n)} ,
\label{III16}
\end{eqnarray}
for the sum of all types of neutrinos as
a function of $\xi^2$ for the first 25 harmonics.
For small $\xi^2$, $\xi^2 \le 0.1$, along with the predominant contribution
of the first two harmonics the contribution of the higher harmonics is significant.
When $\xi^2$ increases, e.g.\ for $\xi^2\simeq 10$, the contribution of higher harmonics
exceeds the contribution of lowest harmonic by orders of magnitude.
The qualitative difference in the relative contribution of higher harmonics
to the total probability of the emission of a photon
(cf.\ Fig.~\ref{Fig:4} in Appendix A~1)
and neutrino pairs is explained by the employed four-fermion structure of the
weak-interaction $e \bar e \nu \bar \nu$ vertex
and the three-particle phase space.
Therefore, the problem of convergence
for the total probability  in case of neutrino emission becomes severe
and deserves special consideration.

\subsection{Overcoming convergence problems}

\begin{figure}[h!]
 \includegraphics[width=0.5\columnwidth]{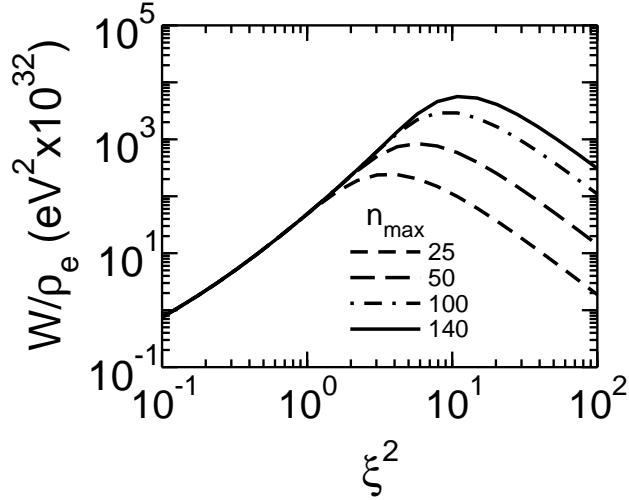}
\caption{\small{
The total probability of neutrino pair emission as
a function of $\xi^2$ for different values of $n_{\rm max}$
from 25 to 140. Kinematics as in Fig.~\ref{Fig:11}.
\label{Fig:12}
}}
\end{figure}

Figure~\ref{Fig:12} illustrates the convergence problem of the total
$\nu \bar \nu$ emission probability with increasing
$\xi^2$ for the case of $\omega_L = 1.55$~eV and $E_e=40$~MeV, where the number
of included harmonics goes up to $n_{\rm max}=140$.
One can see some saturation at $\xi^2\le 2$.
However, for $\xi^2>10$ the difference of probabilities with $n_{\rm max}=25$ and
$n_{\rm max}=140$ is more than two orders of magnitude.

The problem of convergence may be solved by the same method as
proved useful for the one-photon emission (see Appendix \ref{A.1})
using the transformations in Eqs.~(\ref{RitusTransformations_a}) and
(\ref{RitusTransformations_b}). The main difference to the one-photon emission
(cf.~Eq.~(\ref{RitusTransformations2}))
is a modification of the argument of the Airy functions, which emerge from the above
Bessel functions, as
\begin{eqnarray}
y = t \left( 1+\tau^2 + \frac{1+u}{u^2}\frac{M_Q^2}{M_e^2} \right)~,
\label{III17}
\end{eqnarray}
and an additional integration over the invariant mass $dM_Q^2$.
Also, the interference term in Eq.~(\ref{III15}) proportional to the rapidly
oscillating combination $\Phi(y)\Phi'(y)$ is negligible and  can be omitted.
The final expression for the probability $W^{(A)}$ of $\nu \bar\nu $ emission in the
limit of $n_{\rm max}\to \infty$ reads
\begin{eqnarray}
&&W^{{(A)}}(\xi,\chi)=
\frac{\rho_eG_F^2}{48\pi^5}
\int\limits_0^\infty\frac{\sqrt{t}\, du}{(1+u)^2}
\int\limits_0^\infty\,dM_Q^2
\int\limits_{-{\xi}/{2}}^\infty d\tau
\left( F_V^{(A)} C_V^2 + F_A^{(A)} C_A^2 \right)~,
\label{W_asympt}\\
&&F_V^{(A)}=-M_Q^2(2M_e^2 + M_Q^2)\Phi^2(y)
+\frac{2}{t} M_Q^2M_e^2\left(1+\frac{u^2}{2(1+u)}\right)(y\Phi^2(y)+{\Phi'}^2(y))~,
\nonumber\\
&&F^{(A)}_A=M_Q^2(4M_e^2-M_Q^2)\Phi^2(y)
+\frac{2}{t} M_e^2\left(M_Q^2+(M_Q^2+2M_e^2)
\frac{u^2}{2(1+u)}\right)(y\Phi^2(y)+{\Phi'}^2(y))~,
\nonumber
\end{eqnarray}
where $t=(u/2\chi)^{2/3}$.

\begin{figure}[h!]
\vspace{2cm}
 \includegraphics[width=0.5\columnwidth]{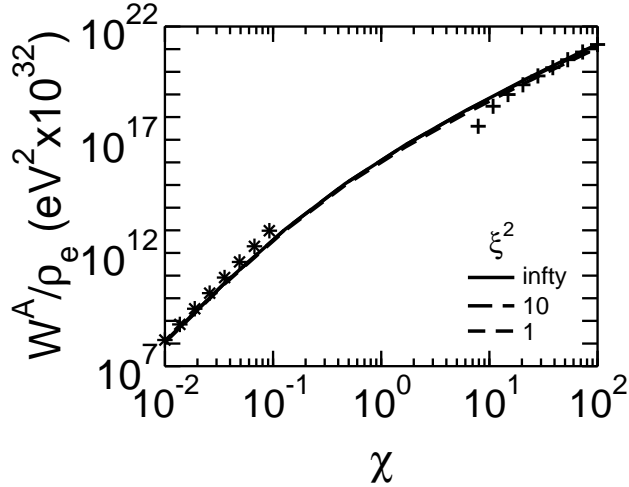}
\caption{\small{
The total probability $W^{(A)}(\xi,\chi)$ of neutrino pair emission as
a function of $\chi$ for different values of $\xi^2$ as indicated in the legend.
The  stars and crosses correspond to the asymptotic
values of $W(\infty, \chi)$ calculated by Eq.~(\ref{III19})
for $\chi\ll 1$ and $\chi\gg1$, respectively.
\label{Fig:13}
}}
\end{figure}

The emission probability $W^{(A)}(\xi,\chi)$ for all three types of neutrinos summed
up is shown in Fig.~\ref{Fig:13} for a
wide region of $\chi$ and $\xi$. The stars and crosses correspond
to the asymptotic values of $W(\infty,\chi)$ for $\chi \ll 1$ and
$\chi\gg 1$, respectively~\cite{Merenkov}:
\begin{eqnarray}
W(\infty,\chi)=
\left\{
\begin{array}{ll}
\frac{\rho_eG_F^2M_e^6\chi^5}{192\sqrt{3}\pi^3}
\left( \frac{49}{6}(C_V^2 +C_A^2) + 63 C_A^2 \right),&\chi\ll 1~,\\
\,&\,\\
\frac{\rho_eG_F^2M_e^6\chi^2}{216\pi^3} (C_V^2 + C_A^2)
\left(\ln{\chi}-0.577 -\frac12\ln{3}-\frac56 \right),&\chi \gg 1~.
\end{array}
\right.
\label{III19}
\end{eqnarray}
Now, the dependence $W^{(A)}(\xi,\chi)$ on $\xi$ is weaker compared with
the photon emission (cf. Fig.~\ref{Fig:6}), mainly because of the additional
integration over $M_Q^2$. Note that the asymptotic estimates
by Eq.~(\ref{III19}) do not match in the intermediate region
$0.1 \lesssim \chi \lesssim 30$.

\begin{figure}[h!]
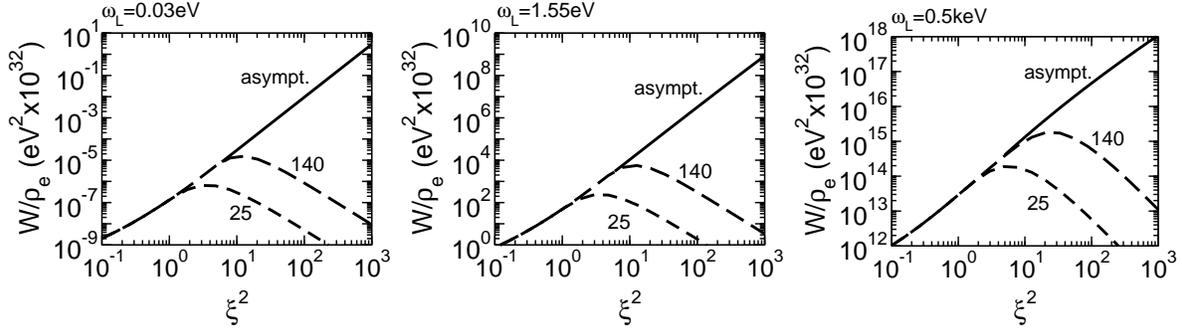

\vspace{2cm}
 \includegraphics[width=0.31\columnwidth]{Fig9a.eps}
 \includegraphics[width=0.31\columnwidth]{Fig9b.eps}
 \includegraphics[width=0.31\columnwidth]{Fig9c.eps}
\caption{\small{
The summed total probability of emission of all three types of neutrinos as
a function of $\xi^2$ for a finite number of harmonics,
$n_{\rm max} = 25$ and 140 (dashed curves),
together with the asymptotic probability given by Eq.~(\ref{W_asympt}) (solid curves).
The electron energy is chosen as  $E_e=40$~MeV, and the laser photon energies $\omega_L$
are 0.03~eV (left panel), 1.55~eV (middle panel) and 0.5~keV (right panel).
\label{Fig:14}
}}
\end{figure}

In Fig.~\ref{Fig:14} we show the total probability $W(\xi,\chi)$
of neutrino pair emission for all neutrinos calculated for
wide initial experimental conditions ranging from
$\omega_L = 0.03$~eV up to $0.5$~keV as a function of $\xi^2$,
where $\xi^2$ varies from 0.1 up to $10^3$. This interval covers possible
experimental conditions illustrated in Fig.~\ref{Fig:1}.
At $\xi^2>6$, $W(\xi,\chi)$ is evaluated using the asymptotic expression
of Eq.~(\ref{W_asympt}). For $\xi^2\le 6$, the probability
might be evaluated as a sum of partial harmonics,
in our case up to $n_{\rm max}=140$.
One can see that, at large $\xi^2$,
the difference between the probability calculated as a sum
of a large but finite number of harmonics and its asymptotic value
which includes an infinite number of harmonics is several orders of magnitude.

Finally, we would like to note the following. In average, the difference of
neutrino pair emission probabilities for $\omega_L=0.03$~eV and 0.5~keV is about
17 orders of magnitude (cf.\ left and right panels in Fig.~\ref{Fig:14}).
This is much greater than the corresponding difference in one-photon emission
shown in the left and right panels of Fig.~\ref{Fig:7}, where the
corresponding difference is about 5 orders of magnitude.
The difference between the two processes
is explained by the different $\chi$ dependence of $W(\xi,\chi)$, shown in
Figs.~\ref{Fig:13} and \ref{Fig:6}, respectively.
The average values of $\chi$ for $\omega_L=0.03$~eV and 0.5~keV
are $10^{-4}$ and $4$, respectively. Therefore, $W(\xi,\chi)$ increases
with $\omega_L$.  The sharp increase
of the probability with $\chi$ in case of neutrino pair emission
is a consequence of the strong energy dependence of the total
probability which can be traced back to the four-fermion structure
of the $ee\nu\bar\nu$ matrix element.

\subsection{Asymmetry of ${\bf \nu_e \bar \nu_e}$ vs.\
${\bf \nu_{\mu, \tau} \bar \nu_{\mu, \tau}}$ emission}

\begin{figure}[h!]
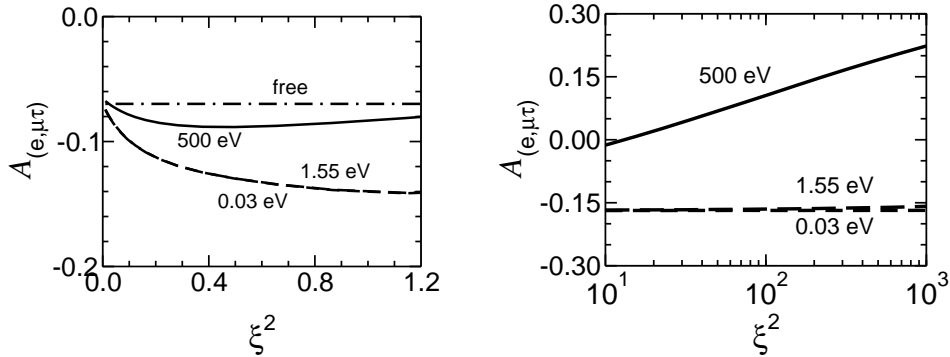

\includegraphics[width=0.35\columnwidth]{Fig10a.eps}\qquad
\includegraphics[width=0.365\columnwidth]{Fig10b.eps}
\caption{\small{
Asymmetry ${\cal A}_{(e,\mu\tau)}$ as a function of $\xi^2$ for different values
of the laser photon energy $\omega_L$. Left panel: low $\xi^2$; probabilities are
calculated as a sum of a finite number of harmonics according to Eq.~(\ref{III11_a}).
The prediction for the $\gamma+e\to e'+\nu\bar\nu$ reaction is shown by the dot-dashed curve
labelled by ''free''.
Right panel: high $\xi^2$; the probabilities
take into account an infinite number of partial harmonics according to Eq.~(\ref{W_asympt}).
\label{Fig:15}
}}
\end{figure}

It seems to be interesting to analyze the asymmetry of
the emission of electron and muon plus tau neutrino pairs
which may be defined as
\begin{eqnarray}
{\cal A}_{(e,\mu\tau)}
=\frac{W_{(e)}-W_{(\mu+\tau)}}
{W_{(e)} + W_{(\mu+\tau)}} .
\label{III20}
\end{eqnarray}
Corresponding predictions for two regions of $\xi^2$ are shown in Fig.~{\ref{Fig:15}}
for different values of the wave field photon energies.
In the region of $\xi^2 \le 1.2$, the probabilities are calculated as a finite sum
of partial harmonics with $n_{\rm max}=50$ (cf.~Eq.~(\ref{III11_a})).
The prediction for the $\gamma+e\to e'+\nu\bar\nu$ reaction
(see Appendix \ref{B})
is shown by the dot-dashed curve
labelled by "free". Predictions for $\omega_L = 0.03$ and 1.55 eV are
virtually identical. For $\xi^2\ll1$, all predictions are close to
the results for the $\gamma+e\to e'+\nu\bar\nu$ reaction:
${\cal A}_{(e,\mu\tau)}\simeq -0.07$.
All asymmetries are small in absolute value and negative,
meaning a slight dominance of the emission of
$\mu + \tau$ neutrinos compared to electron neutrinos.

At large $\xi^2$, the probabilities are calculated as an infinite
number of partial harmonics (cf.~Eq.~(\ref{W_asympt})).
For $\omega_L = 0.03$ and 1.55~eV the range of variation of $\chi$
is $2\cdot 10^{-5}\lesssim\chi\lesssim3\cdot3\cdot10^{-4}$ and
$1.6\cdot 10^{-3}\lesssim\chi\lesssim3\cdot 1.5\cdot10^{-2}$, respectively.
That is, the variable $\chi$ is small and therefore, the asymmetries at these two
energies are close to each other and close to the value
${\cal A}_{(e,\mu\tau)} \simeq -0.17$ which
results from the asymptotic expression Eq.~(\ref{III19}). For
$\omega_L = 0.5$~keV and $0.34\lesssim\chi\lesssim4.8$ the
asymmetry is positive and increases with $\xi^2$.
Note that, within the considered range of $\xi^2$, Eq.~(\ref{III19}) does not apply
and ${\cal A}_{(e,\mu\tau)}$ is smaller
than its asymptotic value $\simeq 0.4$ which would be found from Eq.~(\ref{III19}).
We predict a
distinct dominance of emission of electron neutrino pairs compared to the
sum of muon and tau neutrino pairs
in the keV-range of the photon energy and at large $\xi^2$.
The asymmetry increases as $\ln{\xi^2}$ in the interval of $10\le\xi^2\le1000$.

Finally, we note that our non-perturbative calculation of
${\cal A}_{(e,\mu\tau)}$ is strongly different from the prediction
of pQED (cf. dot-dashed curve in the left panel of Fig.~{\ref{Fig:15}} ) in all
considered intervals of $\omega_L$ and $\xi^2$ unless $\xi^2\lll1$.

\section{Summary}

In summary we have considered $\nu_i\bar\nu_i$
($i=e,\mu,\tau$) emission off an electron
in a strong electromagnetic wave field in a wide
range of energy of the wave field photon energy $\omega_L$
and reduced field intensity $\xi^2$.
Similarly to previous work we expressed the emission
probability as a sum of partial harmonics, where each harmonic
describes the interaction of an electron in-state with $n$ field
photons coherently. For the first time, we made a summation over
harmonics up to a quite large number of harmonics and found that, at large
values of field intensity, $\xi^2>10$, which can be achieved
in current and future laser facilities, the convergence
is rather weak. Therefore, we have elaborated a method allowing for
a complete summation of all partial harmonics.
The method is tested for one-photon emission, i.e.\ non-linear Compton scattering.
Using this new approach we
calculated neutrino pair emission in a region of $\omega_L$
and $\xi^2$ which can be reached experimentally in near future.
We have shown that, at large $\xi^2$, the difference between
the finite and complete sums of partial harmonics reaches a few
orders of magnitudes.

In case of neutrino pair emission we also analyzed the non-trivial
asymmetry between the production of electron and $\mu+\tau$
neutrino pairs. We found that the asymmetry depends strongly
on initial conditions expressed via $\omega_L$ and $\xi^2$. At low $\omega_L$,
the asymmetry is negative corresponding to the dominance of
emission of the $\mu+\tau$ neutrinos, while
at large $\omega_L$ the asymmetry changes the sign indicating the dominance
of electron neutrino pairs.

Finally, we note that all calculations (and conclusions)
have been done for the sake of simplicity for the initial electron energy $E_e=40$~MeV,
which corresponds to the energy of
the superconducting electron accelerator ELBE  in
FZ Dresden-Rossendorf~\cite{ELBE-FZD}. One of the key variable in the
considered processes,
$\chi=\xi\,k \cdot p/M_e^2$, where $k$ and $p$ are the photon and
electron four-momenta, respectively, directly depends on $E_e$.
Therefore, the values of emission probabilities would also depend on $E_e$.
The corresponding evaluation of this dependence as well as
the analysis of other processes will be done in forthcoming papers.
A further step towards realistic estimates is related to the inclusion
of a temporal shape of the external laser wave field,
as considered e.g.\ in \cite{Seipt1,Seipt2,Bulanov2004}.

\acknowledgments
The authors appreciate T.~E. Cowan and D. Seipt
for fruitful discussions.

\appendix

\section{Photon emission off an electron in a strong electromagnetic wave field \label{A}}

The methods employed in Section II are guided by the one-photon emission
process off a Volkov electron, i.e.\ the non-linear Compton effect.
To expose the similarities and differences we recall the essential steps
and clarify further the notation.

\subsection{Strong external field \label{A.1}}

\begin{figure}[h!]
  \includegraphics[width=0.75\columnwidth]{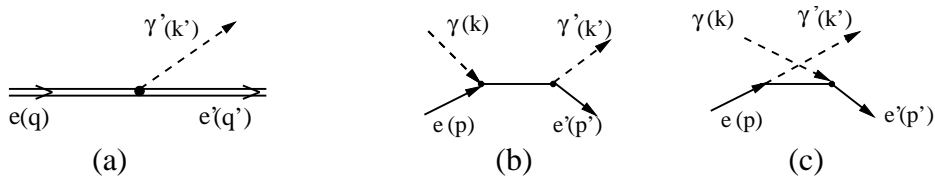}
\caption{\small{
(a)
Diagrammatic representation of the emission of a photon
off an electron in an external wave field. The double lines depict
Volkov states.
(b) and (c) are direct and exchange terms of the
lowest-order diagrams for the perturbative treatment of the Compton
process.
\label{Fig:2} }}
\end{figure}

Let us consider the
emission of a photon off an electron moving in a plane electromagnetic
wave described in the Introduction. 
This process is described by the $S$ matrix element~\cite{LL4}
\begin{eqnarray}
S_{fi}=-ie\int\bar\psi^*_f(\gamma \cdot \varepsilon^*_f)
\psi_i{\rm e}^{ik' \cdot x}\frac{d^4x}{\sqrt{2\omega'}}~,
\label{II9}
\end{eqnarray}
where $\omega'$ is the energy of the emitted photon with four-momentum $k'$ and
$\psi_{i(f)}$ is the electron wave function in the initial (final) state
given by Volkov's solution of the Dirac equation
\begin{eqnarray}
\psi_p=\left[1+\frac{e(\gamma\cdot k)
(\gamma\cdot A)}{2(k\cdot p)}\right]
\,{\exp}\left[
-i\int\limits_0^{k \cdot x}
\frac{e(p\cdot A)}{(k \cdot p)}
\,d\phi'\right]
\frac{u_p}{\sqrt{2q_0}}{\rm e}^{-iq \cdot x}~,
\label{II10}
\end{eqnarray}
where $\gamma$ denote Dirac matrices and $q$ is the quasi-momentum
\begin{eqnarray}
q^\mu=p^\mu-\frac{e^2\langle{A^2}\rangle}{2(k\cdot p)}k^\mu
=p^\mu +\frac{e^2a^2}{2(k\cdot p)}k^\mu
=p^\mu +\frac{\xi^2M_e^2}{2(k\cdot p)}k^\mu
\label{II11}
\end{eqnarray}
of the dressed electron with effective mass
\begin{eqnarray}
q^2 \equiv M_*^2 = M^2_e\left(1-\frac{e^2\langle{A^2}\rangle}{M_e^2}
\right)
=M^2_e\left(1+\xi^2\right)~.
\label{II12}
\end{eqnarray}
Note that Eq.~(\ref{II9}) employs the Furry picture: The field $A^\mu$ is
considered as external classical (background) field, and the emission of a photon
with wave four-momentum $k'$ and polarization $\varepsilon_f$
is described in lowest order of perturbation theory,
see diagram~(a) in Fig.~\ref{Fig:2}.

The dependence of the potential $A$ on $k\cdot x$ in Eq.~(\ref{II10}) results
in the following structure of the $S$  matrix element
\begin{eqnarray}
S_{fi}=\frac{1}{\sqrt{2q_02q_0'2\omega'}}
\int\,M_{fi}(kx)
{\rm e}^{-i(q-q'-k') \cdot x}{d^4x}
=
\sum\limits_{n=-\infty}^{\infty}
\,M^{n}_{fi}(2\pi)^4\delta^4(q + nk-q'-k'),
\label{II13}
\end{eqnarray}
where a Fourier decomposition is used:
\begin{eqnarray}
M(kx,k,k',q,q')=\sum\limits_{n=-\infty}^{\infty}
{\rm e}^{-in\,k \cdot x}\,M^{n}(k,k',q,q')~.
\label{II14}
\end{eqnarray}
Thus, the amplitude is represented as a sum of an infinite number of terms
which are referred to as partial harmonics.
Each harmonic can be attributed to the absorption (emission) of $n$ photons
from (into) the external
field $A$ characterized by the wave four-vector $k$.
For the photon emission off an electron the corresponding conservation law reads
$q+nk=q'+k'$, cf.\ Eq.~(\ref{II13}). In case of on-shell photons,
$k^2={k'}^2=0$ and $k'\cdot q'>0$ hold and therefore $n \ge 1$.
Correspondingly, the differential probability is the infinite sum of partial contributions:
\begin{eqnarray}
dW=\sum\limits_{n=1}^{\infty} dW^{(n)}~,
\label{II15}
\end{eqnarray}
where the partial harmonics $W^{(n)}$ are expressed through
Bessel functions $J_n$ of the first kind~\cite{LL4,NR3}
\begin{eqnarray}
 dW^{(n)}=\frac{\alpha}{2q_0}\,
\frac{du}{(1+u)^2}
\left\{
-2J^2_n(z) +\xi^2(1+\frac{u^2}{2(1+u)})\left(
J^2_{n+1}(z) + J^2_{n-1}(z)-2J^2_n(z)
\right)
\right\}
\label{II16}
\end{eqnarray}
with
\begin{eqnarray}
z=\frac{2n\xi}{\sqrt{1+\xi^2}}
\sqrt{ \frac{u}{u_n} \left(1 - \frac{u}{u_n} \right)}~,
\qquad
u=\frac{k\cdot k'}{k\cdot p'}
=\frac{\omega'(1-\cos\theta)}{E_e'+\omega'\cos\theta}~.
\label{II161}
\end{eqnarray}

\begin{figure}[h!]
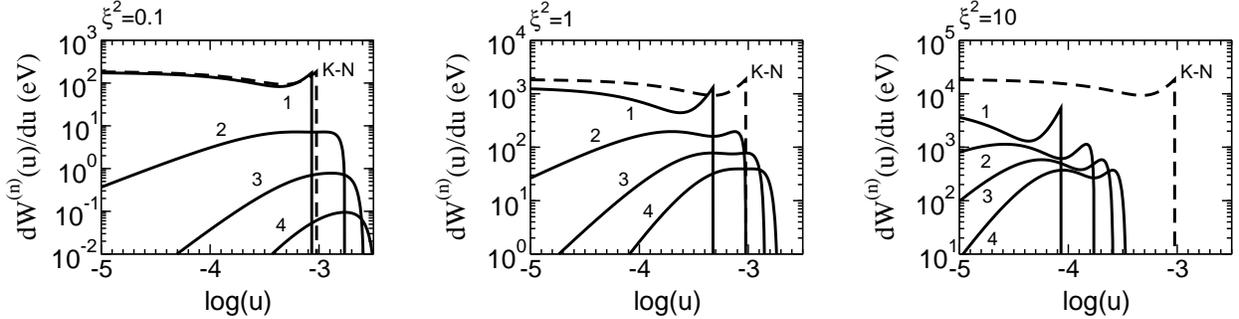

    \includegraphics[width=0.3\columnwidth]{Fig12a.eps}\qquad
    \includegraphics[width=0.3\columnwidth]{Fig12b.eps}\qquad
    \includegraphics[width=0.3\columnwidth]{Fig12c.eps}
   \caption{\small{
   Differential probability of one-photon emission as a function of $\log{u}$
   for $\xi^2=0.1,\,1$ and $10$, shown in left, middle and right panels,
   respectively, for lowest harmonics $n \le 4$.
   The pQCD Compton scattering is shown by the dashed curves labelled by ''K-N''.
   The numbers (from 1 to 4) correspond to the number
   of absorbed photons. The energies of the laser photons
   and incoming electrons are chosen to be $\omega_L = 1.55$~eV and $E_e=40$~MeV,
   respectively.
\label{Fig:3}
}}
  \end{figure}

As an example, we exhibit in Fig.~\ref{Fig:3} the differential probabilities
for photon emission by $40$~MeV electrons colliding head-on with
a $\lambda_L=0.8\,\mu$m ($\omega_L \simeq 1.55\,$MeV) laser beam.
This process can be studied experimentally, e.g.,  at
superconducting electron accelerator ELBE in conjunction with the 150 TW
laser Draco in FZ Dresden-Rossendorf~\cite{ELBE-FZD}. The solid curves correspond
to the partial contributions which come
from the coherent absorption of $n$ photons with $n=1...4$.
The prediction of perturbative QED Compton scattering, described by
the Klein-Nishina formula (see Appendix \ref{A.2}), is shown by the dashed curves.
One can see that for small $\xi^2$, the results for $n=1$ coincide practically
with the result of pQED, cf. also~\cite{Seipt1}.
However, even in this case the contribution of higher
harmonics increases the phase space and modifies the total
probability. When $\xi^2$ increases, the difference between non-perturbative
calculations and pQED is rather large, even for $n=1$. In general, the modification of
the kinematical limit follows Eq.~(\ref{II17}): $u_n$ increases with
$n$ and decreases with $\xi^2$.

\begin{figure}[h!]
 \includegraphics[width=0.5\columnwidth]{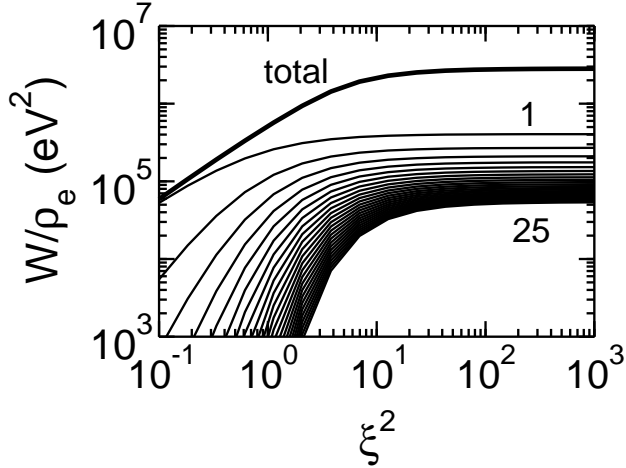}
\caption{\small{
The normalized probability of one-photon emission as
a function of $\xi^2$ for the first 25 harmonics. The thick curve depicts the
completely summed up result. The electron and photon energies are chosen as
$E_e=40$~MeV and $\omega_L= 1.55$~eV, respectively.
\label{Fig:4}
}}
\end{figure}

In Fig.~\ref{Fig:4} we show the total probability of one-photon emission as
a function of $\xi^2$ for the first 25 harmonics. For convenience, we show
reduced probability $W/\rho_e$,
where $\rho_e$ is the density of initial electrons divided
by the electron mass in the electron rest frame.

\begin{figure}[h!]
 \includegraphics[width=0.5\columnwidth]{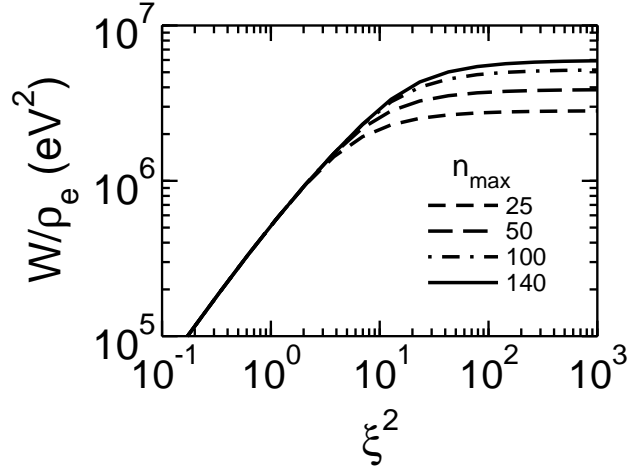}
\caption{\small{
The normalized total probability of one-photon emission as
a function of $\xi^2$ for different values of $n_{\rm max}$
from 25 to 140 as indicated in the legend. Kinematics as in Fig.~\ref{Fig:4}.
\label{Fig:5}
}}
\end{figure}

On can see that for small $\xi^2$, $\xi^2 \ll 1$, the total probability is saturated
by the contribution of the first two
harmonics. However, at large $\xi^2$, $\xi^2 \ge 10$, the contribution of higher harmonics
becomes large and the convergence of the total probability as a function
of the number of harmonics
$n$ is weak. Figure~\ref{Fig:5} illustrates the convergence of
the emission probability with increasing
$\xi^2$ for the case of $\omega_L = 1.55$~eV and $E_e=40$~MeV, where the number
of harmonics increases up to $n_{\rm max}=140$.
One can see some saturation as long as $\xi^2\le 10$. When $\xi^2$ increases,
$n_{\rm max}$ must further increase, in principle, as
$n_{\rm max}\sim \xi^3$~\cite{Ritus79}.

For the calculation of the total probability for large but finite $\xi^2$ we use
the method of Ref.~\cite{Ritus79} based on utilizing the properties of the
Bessel functions $J_n(z)$ at large values of $n$ and $z$ and replacing
the sum of $n$ contributions by an integral over $n$.
Ultimately, this procedure reduces to the following transformations
\begin{eqnarray}
&&J^2_n(z)\rightarrow \frac{1}{\pi^2\xi^2\,t}\Phi^2(y), \label{RitusTransformations_a}\\
&&J^2_{n+1}(z)+J_{n-1}^2(z)-2J^2_n(z) \rightarrow
\frac{2}{\pi^2\xi^4\,t^2}\left(
y\Phi^2(y) -{\Phi'}^2(y)\right)~, \label{RitusTransformations_b}\\
&&dn=\frac{u}{\chi}\,\xi^2\,d\tau~,\qquad -\frac{\xi}{2}\ge\tau < \infty ,
\label{RitusTransformations1}
\end{eqnarray}
where
$\Phi(y)$ is the Airy function, and
\begin{eqnarray}
t = (u/2\chi)^{2/3}~,\qquad y = t(1+\tau^2)~,\qquad \chi = \xi\frac{kp}{M_e^2} .
\label{RitusTransformations2}
\end{eqnarray}

\begin{figure}[h!]
\vspace{2cm}
 \includegraphics[width=0.5\columnwidth]{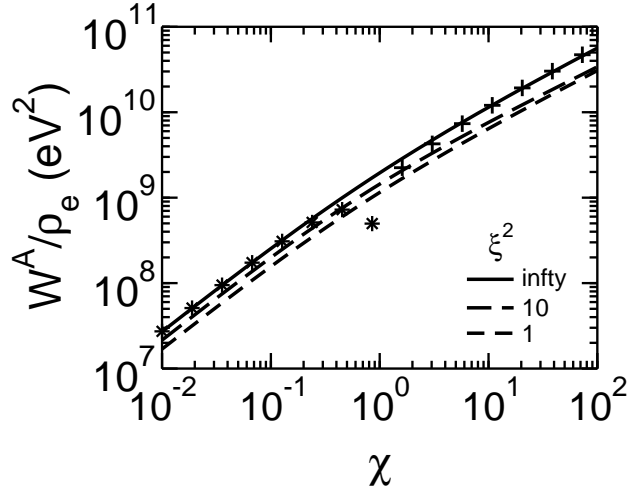}
\caption{\small{
The normalized total probability of one-photon emission $W(\xi,\chi)$ as
a function of $\chi$ for different values of $\xi^2$ as indicated in the legend.
The crosses and stars correspond to the asymptotic
values of $W(\infty, \chi)$ for $\chi\gg 1$ and $\chi\ll1$, respectively.
\label{Fig:6}
}}
\end{figure}

The corresponding probability is expressed in the following form:
\begin{eqnarray}
&&W(\xi,\chi)=\frac{\rho_e \alpha M_e^2}{\pi^2}
\int\limits_0^\infty\frac{\sqrt{t}\,du}{(1+u)^2}
\int\limits_{-\xi/2}^\infty d\tau
\left[-\Phi^2(y) + [1+\frac{u^2}{2(1+u)}]\frac{1}{t}
\left(y\Phi^2(y) - {\Phi'}^2(y)\right)\right]~.\nonumber\\
&&
\label{II18}
\end{eqnarray}
Contrary to Ref.~\cite{Ritus79} and
related papers we do not put $\xi \to \infty$ in the above
integral, which allows to calculate the probabilities at
large but finite values of $\xi^2$ in a wide range of $\chi$.
In Fig.~\ref{Fig:6} we show $W(\xi,\chi)$ as a function of $\chi$
for different values of $\xi^2$.
The crosses and stars correspond to the asymptotic
values of $W(\infty, \chi)$ at $\chi\gg 1$ and $\chi\ll1$,
respectively~\cite{Ritus79}:
\begin{eqnarray}
W(\infty,\chi)=
\left\{
\begin{array}{ll}
\frac{5 \rho_e \alpha M_e^2}{2\sqrt{3}}
\left(1-\frac{8\sqrt{3}}{15}\chi +\frac72\chi^2 + ...\right),&\chi \ll 1,\\
\frac{14 \rho_e \alpha M_e^2}{27}\Gamma(\frac23)(3\chi^{2/3})
\left(
1-\frac{45}{28} \frac{1}{\Gamma(\frac23)(3\chi^{2/3})+ ... }
\right),&\chi \gg 1 .
\end{array}
\right.
\label{II19}
\end{eqnarray}

\begin{figure}[h!]
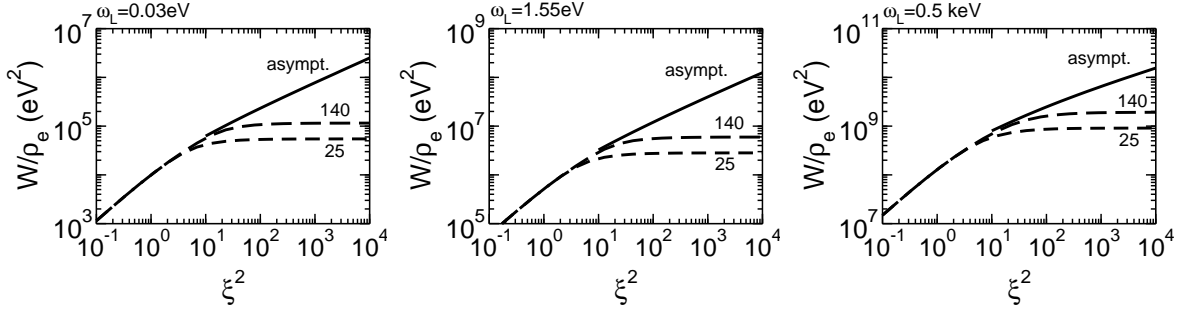

\vspace{2cm}
 \includegraphics[width=0.31\columnwidth]{Fig16a.eps}
 \includegraphics[width=0.31\columnwidth]{Fig16b.eps}
 \includegraphics[width=0.31\columnwidth]{Fig16c.eps}
\caption{\small{
The normalized total probability of one-photon emission as
a function of $\xi^2$ for all harmonics summed (solid curves)
compared to the summed contributions of the first 25 and 140 hamonics
(dashed curves)
with the asymptotic probability given by Eq.~(\ref{II18}).
The electron energy is chosen as  $E_e=40$~MeV, and the photon energies $\omega_L$
are 0.03~eV (left panel), 1.55~eV (middle panel) 0.5~keV (right panel).
\label{Fig:7} }}
\end{figure}

In Fig.~\ref{Fig:7} we show the total probability $W(\xi,\chi)$ calculated for
a wide region of possible experimental conditions ranging from a photon
energy of $\omega_L = 0.03$~eV up $\omega_L = 0.5$~keV as a function of $\xi^2$,
for $\xi^2$ varying from 0.1 up to $10^4$. This interval covers possible
experimental conditions illustrated in Fig.~\ref{Fig:1}.
For $\xi^2>10$, $W(\xi,\chi)$ is evaluated using the asymptotic expression
of Eq.~(\ref{II18}). For $\xi^2\le10$, the probability
is calculated as a sum of partial harmonics, in our case up to $n_{\rm max}=140$.
At the matching point $\xi^2=10$, the difference
between the two expressions is less than 5\%.

\subsection{Perturbative treatment \label{A.2}}

To prove the equivalence of the probabilities defined above to usual
partial cross sections we recollect here also the
perturbative treatment of the scattering of a photon off an electron
(Compton scattering, see middle and right panels in Fig.~\ref{Fig:2}).
The cross section reads in standard notation
\begin{eqnarray}
d \sigma^{\gamma + e' \to \gamma' + e}=\frac{1}{16\pi(s-M_e^2)}|T_{fi}|^2\,dt~,
\label{II1}
\end{eqnarray}
where $s=(p+k)^2$ is the square of total energy and $t=(k-k')^2$ denotes the square of
the momentum transfer. $T_{fi}$ is the invariant amplitude
which consists of direct and crossed terms, schematically depicted
in Fig.~\ref{Fig:2} (b) and (c), respectively,
\begin{eqnarray}
T_{fi}=\,e^2\epsilon^*_\mu(\gamma')
\epsilon_\nu(\gamma) \,
[\bar u(p')M^{\mu\nu}\,u(p)]~,
\label{II2}
\end{eqnarray}
where $\epsilon(\gamma)$ and $\epsilon(\gamma')$ denote the polarization vectors
of incoming ($\gamma$)and outgoing ($\gamma'$) photons, respectively, $u(p)$ and $u(p')$ are
Dirac spinors of incoming and outgoing electrons, normalized
as $\bar uu=2M_e$, and $M^{\mu\nu}$ is the transition operator
\begin{eqnarray}
M^{\mu\nu}=
\gamma^\mu\frac{\gamma\cdot p+\gamma\cdot k +M_e} {2p\cdot k}\gamma^\nu
+
\gamma^\nu\frac{\gamma\cdot k-\gamma\cdot p' +M_e} {2p'\cdot k}\gamma^\mu .
\label{II3}
\end{eqnarray}
In Eq.~(\ref{II1}) averaging over the initial and summation over the final spin states
are provided. The corresponding calculation is described
in text books (for example in~\cite{LL4}).
The result is the Klein-Nishina formula
\begin{eqnarray}
\frac{{d\sigma}}{dt}=\frac{\pi r_0^2M_e^2}{(s-M_e^2)^2}\,F(p,p',k) ,
\label{II4}
\end{eqnarray}
where $r_0\equiv\alpha/M_e$ is the "classical" electron radius and
 \begin{eqnarray}
 F(p,p',k)=
 \left\{
 \left(\frac{M_e^2}{2k\cdot p} + \frac{M_e^2}{2k\cdot p'}\right)^2
+ \left(\frac{M_e^2}{2k\cdot p} + \frac{M_e^2}{2k\cdot p'}\right)
 -\frac14\left(\frac{k\cdot p}{k\cdot p'} + \frac{k\cdot p'}{k\cdot p}\right)
 \right\} .
\label{II5}
\end{eqnarray}
For the further analysis it is convenient to use the
invariant variables $u = \frac{k \cdot k'}{k \cdot p'}$ 
and $dt=2k\cdot p\,du/(1+u)^2$ and to employ the  probability $dW$ of
the $\gamma + e \to \gamma' + e'$ reaction, instead of the cross
section $d\sigma$,
\begin{eqnarray}
dW = \frac{2 s \rho_\gamma}{s + M_e^2}\,d\sigma ,
\label{II7}
\end{eqnarray}
where $\rho_\gamma$ is the photon density (or inverse volume per one photon),
defined similarly to the dependance of $\xi^2$ on $I$ mentioned in the Introduction,
\begin{eqnarray}
\rho_\gamma = \frac{s-M_e^2}{2\sqrt{s}}\frac{M_e^2\xi^2}{4\pi\alpha} .
\label{II8}
\end{eqnarray}
Inspection of Eqs.~(\ref{II4}), (\ref{II7}) and (\ref{II8}) shows that
Compton scattering is independent of $\xi^2$. Variation
of $\xi^2$ (or field intensity $I$)
changes the density of photons $\rho_\gamma$,
and therefore only determines the overall normalization of $dW$.

\section{Perturbative treatment of neutrino pair emission \label{B}}

Consider first neutrino pair production in the reaction
$\gamma + e\to e' + \nu_i\bar \nu_i$. The differential cross section reads
\begin{eqnarray}
\frac{d\sigma^{(i)}}{dM_Q^2 \, du}
=\frac{2}{(8\pi)^3(s-M_e^2)(1+u)^2}\int \frac{d\Omega_\nu}{4\pi}|T^{(i)}|^2 ,
\label{III1}
\end{eqnarray}
where $\Omega_\nu$ is the solid angle
of outgoing neutrino in the $\nu\bar\nu$ rest frame.
The invariant amplitude $T^{(i)}$ in lowest order (tree level)
of the Glashow-Salam-Weinberg model
is described by the Feynman diagrams depicted in
Fig.~4.
Accordingly, the electron neutrino pairs are produced via
exchange of both the charged $W^\pm$ and neutral $Z^0$ vector bosons, while
the muon and tau neutrino pairs are produced only through
the neutral boson exchange when assuming individual lepton number conservation.
In the local limit, where all the momenta involved in the process
are much smaller than the masses of the intermediate vector bosons,
by making use of a Fierz transformation for the charged boson exchange,
one can obtain "universal" effective interactions depicted
in Fig.~\ref{Fig:8b} as direct (a) and exchange (b) terms.
This interaction is described by the effective Lagrangian in Eq.~(\ref{III3}).
Then, the invariant amplitude is expressed as
\begin{eqnarray}
T^{(i)}=\,\frac{eG_F}{\sqrt{2}}\,L^{(i)}_{\alpha}
\epsilon_\beta(\gamma)
\cdot[\bar u(p')M^{(i)\alpha\beta}\,u(p)] ,
\label{III5}
\end{eqnarray}
where $L^{(i)}_\alpha$ is defined by Eq.~(\ref{III6}),
and the transition operator $M^{(i)\mu\nu}$ is defined by Eq.~(\ref{II3}) with
the substitution
\begin{eqnarray}
\gamma^\mu\to\gamma^\mu(C^{(i)}_V - C^{(i)}_A\gamma_5) .
\label{III7}
\end{eqnarray}
Summation over spins of the neutrinos (with the assumption
$m_\nu=0$) and integration over the solid angle in Eq.~(\ref{III1})
leads to
\begin{eqnarray}
\int\frac{d\Omega_\nu}{4\pi}
{\rm Tr}[L_\alpha\, L_\beta^\dagger]
=\frac{8}{3}(Q_\alpha Q_\beta -g_{\alpha\beta} Q^2) ,
\label{III8}
\end{eqnarray}
and one can rewrite Eq.~(\ref{III1}) as
\begin{eqnarray}
\frac{d\sigma^{(i)}}{dM_Q^2 \, du}
&=&\frac{\alpha G_F^2}{512\pi^2(s-M_e^2)(1+u)^2}
F^{(i)}(p,p',k,Q^2)~,
\label{III81}\\
F^{(i)}(p,p',k,Q^2)&=&\frac83
(g^{\alpha\beta}Q^2 - Q^\alpha Q^\beta)
g^{\nu\nu'}\,{\rm Tr}[M^{(i)}_{\alpha\nu}\,{M^{(i)}_{\beta\nu'}}^\dagger] .
\label{III82}
\end{eqnarray}

\end{document}